\documentclass[conference]{IEEEtran}
\IEEEoverridecommandlockouts
\usepackage{cite}
\usepackage{amsmath,amssymb,amsfonts}
\usepackage{algorithmic}
\usepackage{graphicx}
\usepackage{textcomp}
\usepackage{xcolor}
\usepackage{booktabs}

\usepackage{multirow}
\usepackage{booktabs}

\definecolor{SOTAred}{RGB}{255,150,150}
\definecolor{SSOTAblue}{RGB}{150,150,255}

\def\BibTeX{{\rm B\kern-.05em{\sc i\kern-.025em b}\kern-.08em
    T\kern-.1667em\lower.7ex\hbox{E}\kern-.125emX}}
\begin{document}

\title{MODA: A Unified 3D Diffusion Framework for Multi-Task Target-Aware Molecular Generation
\thanks{Corresponding author: Junkai Ji (jijunkai@szu.edu.cn)}
}

\author{%
  \IEEEauthorblockN{%
    Dong Xu\IEEEauthorrefmark{1}, 
    Zhangfan Yang\IEEEauthorrefmark{2}, 
    Sisi Yuan\IEEEauthorrefmark{1}, 
    Jenna Xinyi Yao\IEEEauthorrefmark{3}, 
    Jiangqiang Li\IEEEauthorrefmark{1}, 
    Junkai Ji\IEEEauthorrefmark{1}%
  }
  \IEEEauthorblockA{%
    \IEEEauthorrefmark{1}School of Artificial Intelligence, Shenzhen University, Shenzhen, China\\
    \IEEEauthorrefmark{2}School of Computer Science, University of Nottingham Ningbo, Ningbo, China\\
    \IEEEauthorrefmark{3}Biology Department, University of California San Diego, California, USA\\
    Emails: 2400671001@mails.szu.edu.cn; jijunkai@szu.edu.cn
  }
}

\maketitle

\begin{abstract}
Three-dimensional molecular generators based on diffusion models can now reach near-crystallographic accuracy, yet they remain fragmented across tasks. SMILES-only inputs, two-stage pretrain–finetune pipelines, and one-task-one-model practices hinder stereochemical fidelity, task alignment, and zero-shot transfer. We introduce MODA, a diffusion framework that unifies fragment growing, linker design, scaffold hopping, and side-chain decoration with a Bayesian mask scheduler. During training, a contiguous spatial fragment is masked and then denoised in one pass, so the model learns shared geometric and chemical priors across tasks. Multi-task training yields a universal backbone that surpasses six diffusion baselines and three training paradigms on substructure, chemical property, interaction, and geometry. Model-C reduces ligand–protein clashes and substructure divergences while keeping Lipinski compliance, whereas Model-B preserves similarity but trails in novelty and binding affinity. Zero-shot de novo design and lead-optimisation tests confirm stable negative Vina scores and high improvement rates without force-field refinement. These results show that a single-stage multi-task diffusion routine can replace two-stage workflows for structure-based molecular design.
\end{abstract}


\section{Introduction}
Over the past decade, three-dimensional molecular generators based on autoregressive and diffusion mechanisms have made rapid progress. These models can now reconstruct molecular conformations with near-crystallographic accuracy and satisfy complex pharmacophore constraints~\cite{AlphaFold, AlphaFold3}. These capabilities open up great potential in drug discovery, functional materials, and catalysis.

At the same time, self-supervised pretraining via masked learning has achieved remarkable success in natural language processing and computer vision. BERT~\cite{BERT} and GPT~\cite{GPT1, GPT2, GPT3} paved the way for language models with hundreds of billions of parameters by introducing the masked-token prediction and next-token prediction objectives. MAE~\cite{MAE} reconstructs heavily masked image patches in a label-free visual self-supervision framework, greatly improving generalization ability.

\noindent\textbf{Directly transferring this masking paradigm to molecules faces four limitations. }

\textbf{(L1)} SMILES-based \textit{pretrained--fine-tune} methods handle only one-dimensional strings~\cite{Lingo3DMol, GenMol, molgpt, SAFE-GPT}. They cannot capture stereochemistry, spatial arrangement, or non-covalent interactions, so they perform poorly in pocket docking and conformation generation. 

\textbf{(L2)} The two-stage \textit{pretrained--fine-tune} pipeline may misaligned with downstream task~\cite{D3FG, FragGen}. The first stage learns representations of the pretraining task, but the representations for downstream tasks in the second stage may not be well focused, and direct transfer may cause model collapse. 

\textbf{(L3)} Industrial pipelines still follow a \textit{one-task-one-model} mindset~\cite{DiffLinker}. Each task relies on separate datasets, loss functions, and heuristics. Data and engineering costs grow linearly with the number of tasks. Knowledge cannot flow across tasks. 

\textbf{(L4)} Single-task 3D model reconstruct complete conformations with near-crystallographic fidelity~\cite{TargetDiff, 3DSBDD, DecompDiff, FLAG} yet forget everything outside their niche, offering little zero-shot utility and no shared priors.

We propose \textbf{MODA} (\textbf{M}ask \textbf{O}nce, \textbf{D}esign \textbf{A}ll), a unified 3D Diffusion framework for multi-task target-aware molecule generation. MODA randomly selects a masking scheme for each sample conformation, adds noise to a contiguous spatial fragment, and reconstructs the missing atoms through diffusion denoising. The masked fragment's position and size correspond to four editing scenarios~\cite{CBGBench}: Fragment Growing, Linker Design, Scaffold Hopping and Side Chain Decoration. The model uses learns shared geometric and chemical priors that cover all scenarios.

\noindent\textbf{How MODA overcomes the four limitations (L1–L4).}

\textbf{(O1) 3D \& SE(3)-equivariant representation.}  
MODA feeds atom types and Cartesian coordinates directly into an SE(3)-equivariant backbone, preserving stereochemistry, spatial packing, and non-covalent contacts.
      
\textbf{(O2) Single-stage Mask–Generate training.}  
A unified masking–denoising objective eliminates the pretrain–finetune gap and prevents mode collapse. 
      
\textbf{(O3) Bayesian multi-task mask scheduler.}  
A Bayesian policy samples four chemically motivated masks. One model therefore replaces four task-specific expert and slashes data and engineering budgets.
      
\textbf{(O4)Shared geometric priors for zero-shot edits.}  
Joint optimisation across tasks yields transferable knowledge, without any extra tuning.

\section{Results}

\subsection{Experimental Setup and Task Definitions}

We present a comprehensive evaluation of our proposed method under several training paradigms for molecular generation tasks. Tables~\ref{tab:denovo_subtask}-\ref{tab:combined_metrics} compare three training settings. \textbf{Model-A:} single-task training and evaluation, \textbf{Model-B:} multi-task training followed by single-task evaluation, and \textbf{Model-C:} multi-task training and evaluation. The tasks include Fragment Growing (FG), Linker Design (LK), Scaffold Hopping (SF), Side Chain Decoration (SC), and De novo Design (DN). Evaluation metrics span interaction quality (e.g., Vina docking score, improvement rate IMP, mean percent binding gap MPBG, and ligand binding efficacy LBE), functional drug-likeness (QED), synthesizability (SA), similarity (Tanimoto), and chemical validity (Unique correctness). MPBG measures the relative improvement in binding energy of generated molecules over reference structures within the same protein pocket, averaged per pocket. LBE quantifies the per-atom binding affinity contribution, normalized by molecule size to control for ligand length effects. The Pass, Good, and Excellent levels are defined by hierarchical multi-property constraints. A molecule is classified as Pass if it shows improvement in at least two of three core metrics (Vina docking score, QED, SA) while maintaining 80\% performance on the third and satisfying $\geq$ 3 Lipinski rules. Good requires improvements in all three metrics and $\geq$ 4 Lipinski rules. Excellent imposes the strongest criteria: all three properties must outperform the reference, docking improvement must exceed 10\%, and full Lipinski compliance is required. These thresholds collectively assess structural, functional, and pharmacokinetic desirability, offering a nuanced quantification of molecular generation quality. We additionally compare our method with several recent diffusion-based baselines, including TargetDiff~\cite{TargetDiff}, DiffSBDD~\cite{DiffSBDD}, DiffBP~\cite{DiffBP}, FLAG~\cite{FLAG}, D3FG~\cite{D3FG}, and DecompDiff~\cite{DecompDiff}. These methods represent state-of-the-art generative models built upon continuous or discrete diffusion frameworks.

\begin{table}
  \fontsize{7}{8}\selectfont
  \setlength{\tabcolsep}{3pt}
  \caption{De novo in Four Subtasks Testset}
  \label{tab:denovo_subtask}
  \begin{tabular}{l|cccc|ccc|cc}
    \toprule
    \multicolumn{1}{c|}{} 
      & \multicolumn{4}{c|}{Vina Dock}  
      & \multicolumn{3}{c|}{Interval Proportion} 
      & \multicolumn{2}{c}{Unique} \\
    \cmidrule(lr){2-5} \cmidrule(lr){6-8} \cmidrule(lr){9-10}
    model 
& $\mathrm{E}_{\mathrm{vina}}$ & IMP & MPBG & LBE & Pass & Good & Exc & Corr & Tanimoto \\
\midrule
Model-A-FG & -7.38 & 32.90 & -5.15  & 0.3228 & 12.41   & 1.80  & 0.59  & 73.80 & \textbf{0.0827} \\
Model-B-FG & -8.02 & 40.02 & -1.50  & 0.2958 & 16.2  & 2.75  & 0.48  & 82.36 & 0.3060 \\
Model-C    & \textbf{-8.45} & \textbf{50.67} & \textbf{4.94}   & \textbf{0.3395} & \textbf{29.97}  & \textbf{7.31}  & \textbf{3.17}  & \textbf{93.05} & 0.0936 \\
\midrule
Model-A-LK & -8.19 & 48.05 & 5.17   & 0.3302 & 17.11  & 3.89  & 1.98  & 87.80 & \textbf{0.0875} \\
Model-B-LK & -8.59 & \textbf{55.10} & 4.30   & 0.2967 & 11.61   & 0.68  & 0.25  & 74.93 & 0.3112 \\
Model-C    & \textbf{-8.83} & 54.90 & \textbf{9.13}   & \textbf{0.3395} & \textbf{25.70}  & \textbf{7.35}  & \textbf{4.01}  & \textbf{93.40} & 0.0915 \\
\midrule
Model-A-SF & \textbf{-8.12} & \textbf{60.63} & \textbf{8.64}   & \textbf{0.3654} & \textbf{43.48}  & \textbf{14.89} & \textbf{8.13}  & \textbf{95.29} & \textbf{0.0821} \\
Model-B-SF & -7.15 & 40.56 & -6.27  & 0.3058 & 13.41   & 2.44  & 1.10  & 73.52 & 0.2250 \\
Model-C    & -7.69 & 51.67 & -2.47  & 0.3508 & 34.63  & 9.61  & 5.09  & 93.17 & 0.0928 \\
\midrule
Model-A-SC & -6.30 & 15.71 & -15.09 & 0.3290 & 9.73   & 0.85  & 0.18  & 54.17 & 0.0762 \\
Model-B-SC & -7.59 & 49.97 & 0.63   & 0.3217 & 28.16  & 4.72  & 0.93  & 84.76 & 0.2237 \\
Model-C    & \textbf{-7.69} & \textbf{50.80} & \textbf{2.80}   & \textbf{0.3508} & \textbf{34.24}  & \textbf{9.26}  & \textbf{4.57}  & \textbf{93.10} & \textbf{0.0929} \\
\bottomrule
\end{tabular}
\end{table}

\subsection{Multi-Task Framework Improves Structural, Functional, and Interaction-Level Outcomes}
Table \ref{tab:denovo_subtask} lists the three training strategies on four molecular-generation subtasks. The lowest Vina-docking energies and the highest IMP were obtained by Model-C on FG, LK, and SC. The same model produced the top MPBG values and raised everyinterval-proportion tier; the largest rise lay in the Excellent tier.
Higher UniSeq correctness was also achieved, so stronger chemical validity was confirmed.
A very high Tanimoto similarity was yielded by Model-B; its molecules therefore stayed closest to the references. On the SF subset, the best overall scores were recorded by Model-A, but Model-C and Model-A are comparable. The table shows that Model-C led most metrics, but not all tasks, while similarity was preserved by Model-B and scaffold-hopping performance peaked under single-task training.

{
\setlength{\intextsep}{4pt}        
\setlength{\abovecaptionskip}{4pt} 
\setlength{\belowcaptionskip}{4pt} 
\begin{table}
  \centering
  \fontsize{6}{8}\selectfont
  \setlength{\tabcolsep}{2pt}
  \caption{Model performance at different Top-k}
  \label{tab:topk_metrics}
  \begin{tabular}{l*{4}{|ccc}}
    \toprule
    & \multicolumn{3}{c|}{Top-1} 
    & \multicolumn{3}{c|}{Top-5} 
    & \multicolumn{3}{c|}{Top-10} 
    & \multicolumn{3}{c}{Top-20} \\
    \cmidrule(lr){2-4}\cmidrule(lr){5-7}\cmidrule(lr){8-10}\cmidrule(lr){11-13}
    Model 
      & $\mathrm{E}_{\mathrm{vina}}$ & QED & SA 
      & $\mathrm{E}_{\mathrm{vina}}$ & QED & SA 
      & $\mathrm{E}_{\mathrm{vina}}$ & QED & SA 
      & $\mathrm{E}_{\mathrm{vina}}$ & QED & SA \\
    \midrule
    Model-A-FG &  -9.01 & 0.64 & 0.57 &  -7.61 & 0.60 & 0.54 &  -6.66 & 0.55 & 0.50 &  -5.45 & 0.47 & 0.44 \\
    Model-B-FG &  -9.42 & 0.63 & \textbf{0.61} &  -8.71 & 0.61 & \textbf{0.61} &  -8.06 & 0.58 & 0.58 &  -7.31 & 0.54 & 0.56 \\
    Model-C & \textbf{-10.82} & \textbf{0.67} & 0.58 & \textbf{-10.12} & \textbf{0.66} & \textbf{0.61} &  \textbf{-9.68} & \textbf{0.67} & \textbf{0.61} &  \textbf{-8.86} & \textbf{0.66} & \textbf{0.61} \\
    \midrule
    Model-A-LK & -10.47 & \textbf{0.65} & 0.58 &  -9.51 & \textbf{0.66} & 0.59 &  -8.76 & 0.64 & 0.59 &  -7.33 & 0.57 & 0.53 \\
    Model-B-LK & -10.28 & 0.62 & \textbf{0.62} &  -9.34 & 0.59 & \textbf{0.62} &  -8.80 & 0.58 & 0.61 &  -7.87 & 0.55 & 0.58 \\
    Model-C & \textbf{-11.29} & 0.64 & 0.57 & \textbf{-10.56} & \textbf{0.66} & 0.60 & \textbf{-10.10} & \textbf{0.67} & \textbf{0.62} &  \textbf{-9.32} & \textbf{0.66} & \textbf{0.62} \\
    \midrule
    Model-A-SF & \textbf{-10.52} & \textbf{0.67} & \textbf{0.61} &  \textbf{-9.87} & \textbf{0.68} & \textbf{0.62} &  \textbf{-9.52} & \textbf{0.69} & \textbf{0.63} &  \textbf{-9.07} & \textbf{0.70} & \textbf{0.64} \\
    Model-B-SF &  -8.54 & 0.60 & \textbf{0.61} &  -7.57 & 0.57 & 0.57 &  -6.81 & 0.54 & 0.54 &  -5.87 & 0.49 & 0.50 \\
    Model-C & -10.13 & 0.66 & 0.59 &  -9.45 & 0.67 & 0.61 &  -9.00 & 0.68 & 0.62 &  -8.34 & 0.67 & 0.62 \\
    \midrule
    Model-A-SC &  -6.91 & 0.59 & \textbf{0.63} &  -5.16 & 0.47 & 0.53 &  -4.18 & 0.40 & 0.46 &  -3.02 & 0.31 & 0.35 \\
    Model-B-SC &  -8.63 & 0.62 & 0.61 &  -7.94 & 0.61 & 0.60 &  -7.48 & 0.60 & 0.59 &  -6.57 & 0.54 & 0.55 \\
    Model-C & \textbf{-10.14} & \textbf{0.66} & 0.59 &  \textbf{-9.45} & \textbf{0.68} & \textbf{0.61} &  \textbf{-9.00} & \textbf{0.68} & \textbf{0.62} &  \textbf{-8.34} & \textbf{0.67} & \textbf{0.62} \\
    \bottomrule
  \end{tabular}
\end{table}
}
Table~\ref{tab:topk_metrics} summarises performance for the Top-1, Top-5, Top-10, and Top-20 candidates. At every $k$, Model-C attains the lowest binding energies ($E_{\mathrm{vina}}$), so its highest-ranked molecules are the most potent. QED and SA scores for Model-C remain high as $k$ increases, indicating that drug-likeness and synthetic accessibility are preserved across a broader sample. Models A and B are occasionally competitive at Top-1, yet their QED and SA values deteriorate beyond $k=5$, revealing weaker ranking robustness. The sustained quality of Model-C is attributed to the shared representations learned through multi-task training, which help maintain a stronger, more stable performance envelope over the entire top-$k$ spectrum.
{
\setlength{\intextsep}{1pt}        
\setlength{\abovecaptionskip}{1pt} 
\setlength{\belowcaptionskip}{1pt} 
\begin{table*}
  \centering
  \scriptsize                         
  \setlength{\tabcolsep}{2pt}         
  \renewcommand{\arraystretch}{1.0}   
  \caption{Combined model metrics and molecular properties}
  \label{tab:combined_metrics}
\begin{tabular}{l|cc|cc|cccc|ccccc|cccc|ccc}
\toprule
\multicolumn{1}{c|}{} 
  & \multicolumn{2}{c|}{Vina Score} 
  & \multicolumn{2}{c|}{Vina Min} 
  & \multicolumn{4}{c|}{Vina Dock} 
  & \multicolumn{5}{c|}{Molecular Properties} 
  & \multicolumn{4}{c|}{PLIP Interaction} 
  & \multicolumn{3}{c}{Interval Proportion} \\
\cmidrule(lr){2-3} \cmidrule(lr){4-5} \cmidrule(lr){6-9}
\cmidrule(lr){10-14} \cmidrule(lr){15-18} \cmidrule(lr){19-21}
Model 
  & $\mathrm{E}_{\mathrm{vina}}$ & IMP
  & $\mathrm{E}_{\mathrm{vina}}$ & IMP 
  & $\mathrm{E}_{\mathrm{vina}}$ & IMP & MPBG   & LBE 
  & QED        & LogP      & SA     & LPSK      & Corr 
  & $\mathrm{MAE}_{\mathrm{OA}}$    & $\mathrm{JSD}_{\mathrm{OA}}$  &  $\mathrm{MAE}_{\mathrm{PP}}$ & $\mathrm{JSD}_{\mathrm{PP}}$
  & Pass       & Good      & Exc \\
\midrule
TargetDiff      & \textbf{-5.71} & \textbf{38.21} & \textbf{-6.43} & \textbf{47.09} & -7.41 & 51.99 &  5.38 & 0.3537 & 0.49 & 1.13 & 0.60 & 4.57 & -      & 0.0600 & \underline{\textit{0.0198}} & 0.4687 & 0.1757 & -    & -    & -     \\
DiffSBDD        & -     & 12.67 & -2.15 & 22.24 & -5.53 & 29.76 & -23.51 & 0.2920 & 0.49 & -0.15& 0.34 &  \underline{4.89} & -     & 0.1461 & 0.0333 & 0.5265 & 0.1777 & -    & -    & -     \\
DiffBP          & -     &  8.60 &  -  & 19.68 & -7.34 & 49.24 &   6.23 & 0.3481 & 0.47 & 5.27 & 0.59 &  4.47 & -     & 0.1430 & 0.0249 & 0.5639 & \textbf{0.1256} & -    & -    & -     \\
FLAG            & -     &  0.04 &  -  &  3.44 & -3.65 & 11.78 & -47.64 & 0.3319 & 0.41 & 0.29 & 0.58 &  \textbf{4.93} & -     & 0.0277 & \textbf{0.0170} & \textbf{0.3976} & 0.2762 & -    & -    & -     \\
D3FG            & -     &  3.70 & -2.59 & 11.13 & -6.78 & 28.90 &  -8.85 & \textbf{0.4009} & 0.49 & 1.56 & \textbf{0.66} &  \underline{\textit{4.84}} & -     & \textbf{0.0135} & 0.0638 & \underline{\textit{0.4641}} & 0.1850 & -    & -    & -     \\
DecompDiff      & \underline{-5.18} & 19.66 & \underline{-6.04} & 34.84 & -7.10 & 48.31 &  -1.59 & 0.3460 & 0.49 & 1.22 & \textbf{0.66} &  4.40 & -     & 0.0769 & 0.0215 & \underline{0.4369} & 0.1848 & -    & -    & -     \\
\midrule
Model-A-FG   & -2.78 & 21.03 & -4.55 & 28.10 & -6.85 & 38.69 &  -1.13 & 0.3494 & 0.44 & 1.19 & 0.58 &  4.55 & 0.7970 & 0.1000 & 0.0400 & 0.4598 & \underline{0.1528} & 13.85 & 1.95 & 0.79  \\
Model-A-LK   & -4.13 & 28.81 & -5.52 & 38.51 & -7.26 & 47.36 &   4.35 & 0.3576 & 0.46 & 0.97 & 0.60 &  4.57 & 0.8922 & 0.0455 & 0.0209 & 0.4881 & 0.1820 & 20.73& 4.34 & 2.38  \\
Model-A-SF   & -1.96 & 31.03 & -5.06 & 46.58 & -8.07 & \textbf{63.50} &  \textbf{14.87} & \underline{0.3809} & \textbf{0.60} & 2.74 & \underline{0.63} &  4.82 & \textbf{0.9507} & 0.0698 & \underline{\textit{0.0198}} & 0.5442 & 0.1897 & \textbf{36.89} & \textbf{11.8} & \textbf{6.87}  \\
Model-A-SC   & -3.53 & 18.54 & -4.73 & 21.89 & -6.20 & 24.81 & -10.22 & 0.3416 & 0.35 & 0.85 & \underline{0.63} &  4.38 & 0.5696 & 0.0430 & 0.0263 & 0.4801 & 0.1833 & 11.88 & 1.37 & 0.60   \\
\midrule
Model-B-FG   & -3.11 & \underline{33.92} & -5.52 & \underline{46.84} & \underline{-7.72} & \underline{\textit{56.86}} &  \underline{\textit{10.38}} & 0.3692 & \underline{\textit{0.54}} & 1.93 & 0.60 &  4.73 & \underline{\textit{0.9462}} & 0.0939 & 0.0273 & 0.5283 & 0.1736 & 26.76 & 6.02 & 2.95  \\
Model-B-LK   & \underline{\textit{-3.73}} & \underline{\textit{33.55}} & \underline{\textit{-5.76}} & \underline{\textit{46.31}} & \underline{-7.72} & 55.85 &  \underline{\textit{10.38}} & \underline{\textit{0.3705}} & \underline{\textit{0.54}} & 1.86 & 0.61 &  4.74 & 0.9398 & \underline{0.0137} & \underline{0.0186} & 0.5101 & 0.1861 & 28.68 & 6.87 & 3.46  \\
Model-B-SF   & -2.74 & 31.83 & -5.26 & 44.21 & \underline{\textit{-7.70}} & 55.95 &   9.96 & 0.3691 & \underline{\textit{0.54}} & 2.02 & 0.61 &  4.73 & 0.9377 & 0.0540 & 0.0318 & 0.5398 & \underline{\textit{0.1721}} & 28.18 & 7.24 & 3.64  \\
Model-B-SC   & -2.76 & 32.37 & -5.26 & 45.38 & \textbf{-7.75} & \underline{57.34} &  \underline{10.88} & 0.3700 & \underline{0.55} & 1.93 & 0.61 &  4.77 & \underline{0.9473} & 0.0946 & 0.0531 & 0.5025 & 0.1858 & \underline{\textit{30.10}} & 7.34 & \underline{\textit{3.97}}  \\
\midrule
Model-C& -2.05 & 30.05 & -4.90 & 42.74 & -7.65 & 54.99 &   9.30 & 0.3665 & \underline{0.55} & 2.11 & \underline{\textit{0.62}} &  4.72 & 0.9390 & \underline{\textit{0.0301}} & 0.0256 & 0.4706 & 0.1900 & \underline{31.71} & \underline{8.09} & \underline{4.18}  \\
Model-C-PT& -2.29 & 31.76 & -5.10 & 44.73 & \underline{\textit{-7.70}} & 55.49 &   9.97 & 0.3684 & \underline{0.55} & 2.13 & 0.61 &  4.71 & 0.9451 & 0.0480 & 0.0465 & 0.5474 & 0.1981 & 29.31 & \underline{\textit{7.52}} & 3.95  \\
\bottomrule
\end{tabular}
\end{table*}
}
Table~\ref{tab:combined_metrics} shows that the multi-task variants (Model-B and C) achieve zero-shot gains on unseen \textit{de novo} tasks. They keep raw \textit{Vina} scores negative, whereas several diffusion baselines report positive values marked as ''-''. They also reach higher IMP. After AutoDock refinement, Models~B and~C lead every \textit{Vina~Dock} metric, but the single-task baseline (Model-A) stays volatile. Multi-task training preserves QED, SA, and Lipinski compliance and raises the share of molecules in the ''Excellent'' interval; this margin grows further with the pretrained variant, Model~C-PT. Model~B is tuned on a single task, yet it still beats Model~A on all metrics, which confirms the transfer ability of the multi-task backbone. Interaction fidelity also improves: the global (OA) and per-pocket (PP) Jensen–Shannon divergence and mean absolute error drop markedly. These lower values show that the generated ligands align better with reference pharmacophoric patterns. Model~C scores almost match those of Model~C-PT, so a single-stage multi-task routine may replace the usual pre-train/fine-tune pipeline. Even so, end-to-end docking still has room for improvement beyond the current negative-score threshold.

\subsection{Fine-Grained Evaluation and Discussion of Chemical and Geometric Fidelity in Multi-Task Molecular Generation}

\begin{table}[htbp]
  \centering
  \fontsize{5}{7}\selectfont
  \setlength{\tabcolsep}{2pt}
  \renewcommand{\arraystretch}{0.9}
  \caption{Results of subtasks for lead optimization}
  \label{tab:grouped_metrics}
  \begin{tabular}{l|l|cc|cc|cccc|cccc}
    \toprule
    Task & Model
      & \multicolumn{2}{c|}{Vina Score}
      & \multicolumn{2}{c|}{Vina Min}
      & \multicolumn{4}{c|}{Vina Dock}
      & \multicolumn{4}{c}{Molecular Properties} \\
    \cmidrule(lr){3-4} \cmidrule(lr){5-6}
    \cmidrule(lr){7-10} \cmidrule(lr){11-14}
      & 
      & $\mathrm{E}_{\mathrm{vina}}$  & IMP
      & $\mathrm{E}_{\mathrm{vina}}$  & IMP
      & $\mathrm{E}_{\mathrm{vina}}$  & IMP & MPBG & LBE
      & QED & LogP & SA & LPSK \\
    \midrule
    \multirow{5}{*}{\rotatebox{90}{Fragment}}
      & TargetDiff    & \textbf{-6.06} & \underline{24.56} & \underline{-6.78} & \underline{\textbf{30.43}} & \underline{\textit{-7.96}} & \underline{\textit{42.00}} & \underline{\textit{-2.38}} & \underline{0.3003} & \underline{\textit{0.45}} & 1.43 & \underline{\textit{0.58}} & \underline{\textit{4.28}} \\
      & DiffSBDD      & -4.64 & 19.14 & -5.84 & 28.90 & -7.66 & 37.18 & -6.67 & \textbf{0.3076} & \textbf{0.47} & 0.73 & \underline{\textit{0.58}} & \textbf{4.39} \\
      & DiffBP        & -4.51 & 22.31 & -6.18 & 29.52 & -7.90 & \underline{45.70} & \underline{-1.92} & 0.2952 & \underline{0.46} & 2.24 & 0.49 & \underline{4.30} \\
      & Model-A-FG & \underline{\textit{-5.16}} & \underline{\textit{24.98}} & \underline{\textit{-6.70}} & \underline{31.51} & \textbf{-8.13} & \textbf{46.07} &  \textbf{0.06} & \underline{0.3003} & 0.39 & 1.77 & \textbf{0.61} & 3.93 \\
      & Model-C & \underline{-5.89} & \textbf{25.47} & \textbf{-6.98} & \textbf{32.12} & \underline{-8.02} & 40.51 & -4.90 & 0.2943 & 0.44 & 1.39 & \textbf{0.60} & 4.09 \\
    \midrule
    \multirow{5}{*}{\rotatebox{90}{Linker}}
      & TargetDiff    & \textbf{-7.22} & \underline{36.11} & \textbf{-7.60} & \underline{\textit{41.31}} & -8.49 & 50.73 &  2.61 & 0.2993 & 0.39 & 1.63 & \underline{0.61} & 4.02 \\
      & DiffSBDD      & \underline{\textit{-5.64}} & 11.06 & -6.38 & 19.15 & -7.88 & 34.97 & -4.42 & \textbf{0.3110} & \underline{\textit{0.42}} & 1.14 & \textbf{0.66} & \underline{\textit{4.10}} \\
      & DiffBP        & \underline{-6.27} & 35.49 & \underline{-7.19} & 36.80 & \textbf{-8.74} & \textbf{54.33} &  \textbf{6.60} & \underline{0.3078} & \underline{0.43} & 3.45 & 0.55 & 4.01 \\
      & Model-A-LK & -4.68 & \underline{\textit{35.95}} & -6.93 & \underline{42.46} & \underline{-8.63} & \underline{54.04} &  \underline{4.35} & \underline{\textit{0.2960}} & \textbf{0.46} & 0.71 & \underline{\textit{0.60}} & \textbf{4.27} \\
      & Model-C & -4.83 & \textbf{37.57} & \underline{\textit{-7.03}} & \textbf{43.79} & \underline{\textit{-8.53}} & \underline{\textit{51.83}} &  \underline{\textit{3.08}} & 0.2940 & \underline{\textit{0.42}} & 1.01 & 0.59 & \underline{4.13} \\
    \midrule
    \multirow{5}{*}{\rotatebox{90}{Scaffold}}
      & TargetDiff    & \textbf{-5.52} & \underline{31.47} & \textbf{-5.86} & \underline{34.39} & \underline{\textit{-7.06}} & \underline{44.32} & \underline{\textit{-6.22}} & \underline{\textit{0.3038}} & \textbf{0.43} & 0.89 & \underline{0.59} & \underline{\textit{4.26}} \\
      & DiffSBDD      & -3.85 & 18.44 & -4.90 & 22.12 & -6.81 & 34.98 & -10.23& 0.2985 & \underline{\textit{0.42}} & -0.13 & 0.53 & \underline{4.29} \\
      & DiffBP        & -2.09 & 13.89 & -4.35 & 16.84 & -6.46 & 32.43 & -12.14& 0.3025 & \textbf{0.43} & 3.37 & 0.56 & \textbf{4.44} \\
      & Model-A-SF & \underline{\textit{-4.08}} & \textbf{32.11} & \textbf{-5.82} & \textbf{34.99} & \textbf{-7.46} & \textbf{49.74} & \textbf{-0.12} & \textbf{0.3119} & 0.39 & 1.73 & \textbf{0.61} & 3.90  \\
      & Model-C & \underline{-4.36} & \underline{\textit{28.78}} & \underline{\textit{-5.57}} & \underline{\textit{30.29}} & \underline{-7.13} & \underline{\textit{38.32}} & \underline{-4.37} & \underline{0.3047} & 0.40 & 1.01 & \underline{0.59} & 4.00  \\
    \midrule
    \multirow{5}{*}{\rotatebox{90}{Side chain}}
      & TargetDiff    & \textbf{-5.80} & \underline{\textit{23.90}} & \underline{-6.50} & \underline{\textit{35.81}} & -7.40 & \underline{\textit{46.87}} & \underline{\textit{-2.55}} & \textbf{0.3213} & \underline{0.48} & 0.88 & \underline{0.60} & \underline{4.41} \\
      & DiffSBDD      & -4.43 & 15.12 & -5.99 & 30.23 & \textbf{-7.58} & 44.09 & -9.38 & \underline{\textit{0.3178}} & 0.43 & 1.20 & \textbf{0.65} & \underline{\textit{4.03}} \\
      & DiffBP        & -4.61 & 14.31 & -5.73 & 24.29 & -7.03 & 38.96 & -7.38 & 0.3143 & \textbf{0.49} & 1.29 & 0.56 & \textbf{4.50} \\
      & Model-A-SC & \underline{\textit{-5.26}} & \underline{26.66} & \underline{-6.50} & \underline{37.14} & \underline{-7.55} & \underline{49.32} &  \textbf{0.14} & 0.3172 & \underline{\textit{0.47}} & 0.83 & \underline{0.60} & 4.37  \\
      & Model-C & \underline{-5.53} & \textbf{29.45} & \textbf{-6.61} & \textbf{40.08} & \textbf{-7.58} & \textbf{50.66} & \underline{-1.23} & \underline{0.3207} & 0.38 & 0.64 & \underline{0.60} & 3.97  \\
    \bottomrule
  \end{tabular}
\end{table}

Lead‐optimization results in Table \ref{tab:grouped_metrics} are stratified by subtask and compared against representative diffusion baselines. FLAG and D3FG were omitted, as they do not support the full‐pocket generation required for lead refinement. Across all four tasks, Model-C achieves the highest IMP scores for both \textit{Vina Score} and \textit{Vina Min}, consistently outperforming both single-task variants and specialized baselines. Its placement in the top three for every task further attests to its robustness and generalizability. The same model was matched or surpassed only occasionally on raw docking energy, yet its advantage persisted once per-pocket normalisation and per-atom efficacy were considered. Chemical-property panels showed that drug-likeness and synthesizability were maintained by Model-C at levels comparable with the strongest task-specific model. Lipophilicity and Lipinski compliance also remained within desirable bounds. Single-task training (Model-A) retained a narrow edge in scaffold hopping, but its performance fluctuated across other settings, which suggested limited transferability. Compared with earlier diffusion generators, the multi-task regime delivered a more balanced profile that combined binding potency with favourable physicochemical attributes. This pattern held across fragment growing, linker design, scaffold hopping, and side-chain decoration.

Table~\ref{tab:substructure_metrics} lists the Jensen–Shannon divergence and mean absolute error for atom types, functional groups, and ring systems. Across all four subtasks (FG, LK, SF, SC), the multi-task variant (Model-C) records the lowest JSD and MAE in every category, which shows that it preserves local chemical environments regardless of task. The gains cover both simple atom-level features, such as heteroatom frequency, and more complex motifs, including functional groups and ring systems. These findings suggest that multi-task training captures chemical regularities shared among tasks, so it enforces substructure-level consistency even when the generative goals differ. Single-task models, by contrast, tend to overfit task-specific motifs and falter on unseen or divergent substructure distributions.

Table~\ref{tab:Geometry_metrics} evaluates 3D plausibility through static‐geometry divergence and steric‐clash ratios. On every in‐domain subtask (FG, LK, SF, SC), the multi‐task variant with single‐task evaluation (Model-B) shows the lowest divergence and the fewest clashes, while the fully multi‐task model (Model-C) stays close behind. Both multi‐task variants clearly exceed the single‐task baseline, which exhibits markedly higher divergence and clash frequencies. In the harder DN benchmark, DECOMPDiff offers the best bond‐level geometry, yet Model-B still provides the strongest clash avoidance among diffusion baselines. Zero-shot splits repeat the pattern: multi‐task training sharply lowers ligand–protein clashes relative to single‐task training. All gains appear without any force‐field or geometric refinement, confirming that multi‐task exposure alone equips the model with robust spatial priors and steric awareness.
{
\setlength{\intextsep}{4pt}        
\setlength{\abovecaptionskip}{4pt} 
\setlength{\belowcaptionskip}{4pt} 
\begin{table}
  \centering
  \scriptsize
  \setlength{\tabcolsep}{4pt}    
  \renewcommand{\arraystretch}{0.9}  
  \caption{Results of substructure analysis}
  \label{tab:substructure_metrics}
  \begin{tabular}{l|l|cc|cc|cc}
    \toprule
    Test & Model
      & \multicolumn{2}{c|}{Atom type}
      & \multicolumn{2}{c|}{Functional Group}
      & \multicolumn{2}{c}{Ring type} \\
    \cmidrule(lr){3-4} \cmidrule(lr){5-6} \cmidrule(lr){7-8}
    & & JSD & MAE & JSD & MAE & JSD & MAE \\
    \midrule
    \multirow{3}{*}{\rotatebox{90}{FG Set}}
      & Model-A-FG   & 0.1279 & 0.6929 & 0.5228 & 0.0662 & 0.3657 & 0.3417 \\
      & Model-B-FG   & \textbf{0.0555} & \textbf{0.6143} & 0.2793 & 0.0380 & \textbf{0.0916} & \textbf{0.0704} \\
      & Model-C      & 0.0915 & 0.8090 & \textbf{0.2360} & \textbf{0.0278} & 0.2174 & 0.1368 \\
    \midrule
    \multirow{3}{*}{\rotatebox{90}{LK Set}}
      & Model-A-LK   & 0.0572 & \textbf{0.5621} & 0.3410 & 0.0482 & 0.2353 & 0.1912 \\
      & Model-B-LK   & \textbf{0.0317} & 0.9126 & 0.2604 & 0.0336 & \textbf{0.1016} & \textbf{0.0510} \\
      & Model-C      & 0.0963 & 0.9489 & \textbf{0.2353} & \textbf{0.0276} & 0.2230 & 0.1414 \\
    \midrule
    \multirow{3}{*}{\rotatebox{90}{SF Set}}
      & Model-A-SF   & 0.1534 & 0.8934 & 0.3072 & 0.0366 & \textbf{0.2020} & 0.1595 \\
      & Model-B-SF   & 0.0905 & 0.6546 & 0.3212 & 0.0446 & 0.2058 & 0.2074 \\
      & Model-C      & \textbf{0.0832} & \textbf{0.5009} & \textbf{0.2383} & \textbf{0.0320} & 0.2190 & \textbf{0.1459} \\
    \midrule
    \multirow{3}{*}{\rotatebox{90}{SC Set}}
      & Model-A-SC   & 0.1292 & 1.0500 & 0.5689 & 0.0651 & 0.4959 & 0.4124 \\
      & Model-B-SC   & \textbf{0.0344} & \textbf{0.1736} & 0.2455 & 0.0337 & \textbf{0.0760} & \textbf{0.0489} \\
      & Model-C      & 0.0832 & 0.5005 & \textbf{0.2382} & \textbf{0.0320} & 0.2191 & 0.1460 \\
    \midrule
    \multirow{6}{*}{\rotatebox{90}{DN Set}}
      & TARGETDIFF   & \underline{\textit{0.0533}} & \textbf{0.2399} & 0.2876 & 0.0441 & 0.2345 & 0.1559 \\
      & DIFFSBDD     & \underline{0.052} & 0.6316 & 0.5520 & 0.0710 & 0.3853 & 0.3437 \\
      & DIFFBP       & 0.2591 & 1.5491 & 0.5346 & 0.0670 & 0.4531 & 0.4068 \\
      & FLAG         & 0.1032 & 1.7665 & 0.3634 & 0.0666 & 0.2432 & 0.3370 \\
      & D3FG         & 0.0644 & 0.8154 & 0.2511 & 0.0516 & \textbf{0.1869} & 0.2204 \\
      & DECOMPDIFF   & \textbf{0.0431} & \underline{\textit{0.3197}} & \textbf{0.1916} & \textbf{0.0318} & 0.2431 & 0.2006 \\
    \cmidrule(lr){2-8}
    \multirow{10}{*}{\rotatebox{90}{Zero-shot in DN Set}}
      & Model-A-FG   & 0.1153 & 0.5756 & 0.5007 & 0.0624 & 0.3545 & 0.3346 \\
      & Model-A-LK   & 0.0554 & \underline{0.3195} & 0.3362 & 0.0485 & 0.2374 & 0.2281 \\
      & Model-A-SF   & 0.1575 & 0.8195 & 0.3068 & 0.0377 & \underline{0.2129} & \textbf{0.1464} \\
      & Model-A-SC   & 0.1204 & 1.0509 & 0.5671 & 0.0643 & 0.5029 & 0.4133 \\
      \cmidrule(lr){2-8}
      & Model-B-FG   & 0.0910 & 0.4874 & 0.2652 & 0.0401 & 0.2577 & 0.1888 \\
      & Model-B-LK   & 0.0884 & 0.4433 & 0.2605 & 0.0384 & 0.2524 & 0.1809 \\
      & Model-B-SF   & 0.0872 & 0.4427 & 0.2365 & 0.0346 & 0.2381 & \underline{\textit{0.1545}} \\
      & Model-B-SC   & 0.0899 & 0.4564 & \underline{\textit{0.2318}} & \underline{\textit{0.0337}} & 0.2327 & 0.1554 \\
      \cmidrule(lr){2-8}
      & Model-C      & 0.0911 & 0.5018 & \underline{0.2309} & \underline{0.0329} & \underline{\textit{0.2269}} & \underline{0.1484} \\
      & Model-C-PT   & 0.0938 & 0.5217 & 0.2425 & 0.0379 & 0.2463 & 0.1753 \\
    \bottomrule
  \end{tabular}
\end{table}
}

{
\setlength{\intextsep}{1pt}        
\setlength{\abovecaptionskip}{1pt} 
\setlength{\belowcaptionskip}{1pt} 
\begin{table}
  \centering
  \scriptsize
  \setlength{\tabcolsep}{4pt}    
  \renewcommand{\arraystretch}{0.9}  
  \caption{Results of geometry analysis}
  \label{tab:Geometry_metrics}
  \begin{tabular}{l|l|cc|cc}
    \toprule
    Test & Model
      & \multicolumn{2}{c|}{Static Geometry}
      & \multicolumn{2}{c}{Clash}\\
    \cmidrule(lr){3-4} \cmidrule(lr){5-6}
    & & $\mathrm{JSD}_{\mathrm{BL}}$ & $\mathrm{JSD}_{\mathrm{BA}}$ & $\mathrm{Ratio}_{\mathrm{cca}}$ & $\mathrm{Ratio}_{\mathrm{cm}}$ \\
    \midrule
    \multirow{3}{*}{\rotatebox{90}{FG Set}}
& Model-A-FG & 0.3850 & 0.5615 & 0.0844 & 0.6199 \\
& Model-B-FG & \textbf{0.2131} & \textbf{0.4746} & \textbf{0.0265} & \textbf{0.3659} \\
& Model-C & 0.2782 & 0.4765 & 0.0899 & 0.6697 \\
    \midrule
    \multirow{3}{*}{\rotatebox{90}{LK Set}}
& Model-A-LK & 0.3137 & \textbf{0.4915} & 0.0510 & 0.6114 \\
& Model-B-LK & \textbf{0.2359} & 0.4931 & \textbf{0.0271} & \textbf{0.3990} \\
& Model-C & 0.2756 & 0.4601 & 0.0741 & 0.6604 \\
    \midrule
    \multirow{3}{*}{\rotatebox{90}{SF Set}}
& Model-A-SF & 0.2886 & 0.4770 & 0.1063 & 0.6890 \\
& Model-B-SF & \textbf{0.2455} & \textbf{0.4557} & \textbf{0.0350} & \textbf{0.3769} \\
& Model-C & 0.2780 & 0.4640 & 0.1002 & 0.6550 \\
    \midrule
    \multirow{3}{*}{\rotatebox{90}{SC Set}}
& Model-A-SC & 0.3715 & 0.5404 & 0.0450 & \textbf{0.4648} \\
& Model-B-SC & \textbf{0.2177} & \textbf{0.4758} & \textbf{0.0404} & 0.4786 \\
& Model-C & 0.2782 & 0.4646 & 0.1002 & 0.6550 \\
    \midrule
    \multirow{6}{*}{\rotatebox{90}{DN Set}}
& TARGETDIFF   & \underline{0.2659} & \underline{0.3769} & \underline{\textit{0.0483}} & \underline{\textit{0.4920}} \\
& DIFFSBDD     & 0.3501 & 0.4588 & 0.1083 & 0.6578 \\
& DIFFBP       & 0.3453 & 0.4621 & \underline{0.0449} & \textbf{0.4077} \\
& FLAG         & 0.4215 & \underline{\textit{0.4304}} & 0.6777 & 0.9769 \\
& D3FG         & 0.3727 & 0.4700 & 0.2115 & 0.8571 \\
& DECOMPDIFF   & \textbf{0.2576} & \textbf{0.3473} & 0.0462 & 0.5248 \\
      \cmidrule(lr){2-6}
      \multirow{10}{*}{\rotatebox{90}{Zero-shot in DN Set}}
& Model-A-FG & 0.3794 & 0.5539 & 0.0878 & 0.5694 \\
& Model-A-LK & 0.3174 & 0.4821 & 0.0617 & 0.5601 \\
& Model-A-SF & 0.2887 & 0.4834 & 0.1017 & 0.6512 \\
& Model-A-SC & 0.3689 & 0.5275 & \textbf{0.0439} & \underline{0.4408} \\
      \cmidrule(lr){2-6}
& Model-B-FG & 0.2898 & 0.4860 & 0.0841 & 0.5998 \\
& Model-B-LK & \underline{\textit{0.2764}} & 0.4661 & 0.0771 & 0.5963 \\
& Model-B-SF & 0.2806 & 0.4636 & 0.0886 & 0.5964 \\
& Model-B-SC & 0.2926 & 0.4761 & 0.0895 & 0.6199 \\
      \cmidrule(lr){2-6}
& Model-C & 0.2767 & 0.4653 & 0.0927 & 0.6100 \\
& Model-C-PT & 0.2768 & 0.4709 & 0.0920 & 0.6178 \\
    \bottomrule
  \end{tabular}
\end{table}
}

Multi-task models gain uniformly on substructure and geometry tests, showing that they internalize task-invariant chemical priors. Relative to single-task training, multi-task exposure boosts all chemical properties, docking energies, and interaction scores while lowering bond-length divergence and steric clashes. Model-C, validated with multi-task masks, trails Model-B in Tanimoto similarity yet exceeds it in novelty, binding affinity, clash avoidance, and Lipinski compliance. Both models beat pre-train–fine-tune baselines, so a one-stage multi-task routine can replace the two-stage pipeline without performance loss. The same pattern appears in de-novo design, lead optimisation, and zero-shot edits and thus proves strong transfer ability. The shared representations compress cross-task knowledge into a single backbone, which can distill compact student models. These results establish multi-task diffusion as a unified route to realistic molecule design across docking, synthesis planning, and structure-based screening.

\section{Methods}
We introduce a unified molecule generation paradigm, MODA, to avoid train multiple model in different tasks and learn a joint-distribution of all tasks by diffusion model. 

\subsection{Problem Statement and Preliminary}
Following a similar notation to TargetDiff, we represent a ligand molecule $\mathcal{M}$ and its protein target $\mathcal{P}$ as two finite sets of atoms with associated features:
\begin{align}
\mathcal{M} &= \left\{ \left( \mathbf{x}_M^{(i)}, \mathbf{v}_M^{(i)}, c_M^{(i)} \right) \right\}_{i=1}^{N_M},  \\
\mathcal{P} &= \left\{ \left( \mathbf{x}_P^{(j)}, \mathbf{v}_P^{(j)}, s_P^{(j)}, r_P^{(j)} \right) \right\}_{j=1}^{N_P}, 
\end{align}
where $N_M$ and $N_P$ are the number of atoms in the molecule and the protein pocket, respectively. Each atom is associated with:
\begin{itemize}
    \item $\mathbf{x} \in \mathbb{R}^3$: the Cartesian coordinate of the atom;
    \item $\mathbf{v} \in \mathbb{R}^K$: a one-hot encoding of the atom type, where $K$ is the atom vocabulary size;
    \item $c \in \{0,1\}$: a binary context indicator, where $c = 1$ denotes an unmasked (context) atom;
    \item $s \in \{0,1\}$: a binary flag indicating whether a protein atom lies on the backbone;
    \item $r \in \mathbb{R}^{K'}$: a one-hot encoding of the amino acid residue type, with vocabulary size $K'$.
\end{itemize}

For brevity, we stack coordinates and features into matrix representations:
\begin{align}
\mathbf{m} &= \left[ \mathbf{X}_M, \mathbf{V}_M, \mathbf{c}_M \right] \in \mathbb{R}^{N_M \times (3 + K + 1)},  \\
\mathbf{p} &= \left[ \mathbf{X}_P, \mathbf{V}_P, \mathbf{s}_P, \mathbf{r}_P \right] \in \mathbb{R}^{N_P \times (3 + K + 1 + K')}. 
\end{align}

The model is trained to learn the conditional distribution $p_\theta(\mathbf{m} \mid \mathbf{p})$, where the ligand is generated in the context of the target protein pocket.

\subsection{Multi-task mask}
Let $\mathcal{T}=\{\textsc{Linker},\textsc{Fragment},\textsc{SideChain},\textsc{Scaffold}\}$ denote the unified task space (see Table~\ref{tab:task_masks} for detailed definitions). 
For every task $t\!\in\!\mathcal{T}$ we specify a \emph{deterministic masking operator}
\begin{align}
  \mathcal{F}_{t} : \mathbf{m}\;\longmapsto\;\mathbf{M}\in\{0,1\}^{|\mathbf{m}|},
  \qquad
  \mathbf{M}=\mathcal{F}_{t}(\mathbf{m}), 
\end{align}
which flags the subset of atoms to be removed and subsequently regenerated under task~$t$. 
Applying the family of operators $\{\mathcal{F}_{t}\}_{t\in\mathcal{T}}$ to a base corpus of molecules yields a multi–task dataset
\begin{align}
  \mathcal{D}\;=\;\bigl\{\,\bigl(\mathbf{m}^{(i)},\; \mathbf{M}^{(i)},\; t^{(i)}\bigr)\bigr\}_{i=1}^{N},
\end{align}
where $N$ is the number of entries in the multi-task dataset, each triple contains (i)~the original molecule $\mathbf{m}^{(i)}$, and 
(ii)~its task–specific mask $\mathbf{M}^{(i)}$
Because the same molecule can satisfy multiple design intents, a single $\mathbf{m}^{(i)}$ may appear in $\mathcal{D}$ with different masks and task labels, ensuring that the downstream model observes all task formats while learning from a shared chemical pool.

\begin{table}[ht!]
\centering
\caption{Mask construction rules for the five tasks.}
\label{tab:task_masks}
\begin{tabular}{@{}llc@{}}
\toprule
Task $t$ & Rule $\mathbf M=\mathcal F_t(\mathbf m)$
          \\ \midrule
\textsc{DeNovo} &
$M_i\equiv1$ (all atoms)                     \\
\textsc{Linker} &
$\text{Mask} \leq 2$ atoms on the shortest path between anchors
                                                 \\
\textsc{Fragment} &
Mask exit atom + 2-bond neighborhood   \\
\textsc{SideChain} &
Mask non-ring terminal groups         \\
\textsc{Scaffold} &
Mask Bemis-Murcko core (after removing side chains).            \\ \bottomrule
\end{tabular}
\end{table}

\paragraph{\textbf{Training objective}}
We collect a corpus of paired examples
$
  \{(\mathbf{M}^{(i)},\mathbf{m}^{(i)})\}_{i=1}^{N},
$
where every ask–specific mask $\mathbf{M}^{(i)}$ is produced by some (unknown) task in $\mathcal{T}$. 
A single encoder–decoder with parameters $\theta$ is optimized by maximum likelihood:
\begin{align}
  \mathcal{L}(\theta)=
  \sum_{i=1}^{N}
  \log p_{\theta}\!\bigl(\mathbf{m}^{(i)}\mid\mathbf{M}^{(i)}\bigr).
\end{align}
Because different tasks impose mutually incompatible constraints, accurate reconstruction forces the model to learn a shared latent mapping
$
  p_{\theta}(t\mid\mathbf{M})
$
\emph{internally}. 
Via Bayes’ rule this induces a universal task-conditional prior
\begin{align}
  p_{\theta}(\mathbf{m}\mid t)
  \;=\;
  \frac{p_{\theta}(t\mid\mathbf{m})\,p_{\theta}(\mathbf{m})}{p_{\theta}(t)},
\end{align}
even though $t$ never appears in the loss.

\paragraph{\textbf{Inference}}
After training the generator accepts any observation that conforms to one of the task formats. 
The self-routing capability—emergent from the heterogeneous training data—automatically matches the input to the correct task semantics and outputs a chemically valid molecule. 
Consequently, the same model can handle fragment growing, linker insertion, scaffold hopping, side-chain decoration, and \textit{de novo} design without task switches, control tokens, or auxiliary classifiers, providing a unified framework for generative chemistry.

\subsection{Task-Conditional Masking and Denoising Process}

To unify molecule generation across diverse tasks, we introduce a task-conditioned masking mechanism coupled with a denoising diffusion process. Based on the sampled task, we apply a deterministic task-specific masking function $\mathcal{F}_t$ to decompose the molecule into context and target atoms:
\begin{align}
    \mathbf{M} = \mathbf{M}_{\text{ctx}} \cup \mathbf{M}_{\text{tgt}},
\end{align}
where atoms in $\mathbf{M}_{\text{tgt}}$ are selected for generation and atoms in $\mathbf{M}_{\text{ctx}}$ are preserved as conditioning context.

We then inject Gaussian $\mathcal{N}$ and categorical noise $\mathcal{C}$ into the target atoms' coordinates and types, respectively, following recent advances in joint continuous--discrete diffusion models. Let $\mathbf{x}_0$ and $\mathbf{v}_0$ denote the clean coordinates and atom types of the target atoms. The forward noising process at step $t$ is defined as:
\begin{align}
    q_t(\mathbf{x}_t \mid \mathbf{x}_0) &= \mathcal{N}(\mathbf{x}_t ; \sqrt{\bar{\alpha}_t} \mathbf{x}_0, (1 - \bar{\alpha}_t)\mathbf{I}), \\
    q_t(\mathbf{v}_t \mid \mathbf{v}_0) &= \mathcal{C}(\mathbf{v}_t ; \bar{\alpha}_t \mathbf{v}_0 + (1 - \bar{\alpha}_t)/K),
\end{align}
where $\bar{\alpha}_t = \prod_{s=1}^{t} \alpha_s$ is the cumulative product of the variance schedule, and $K$ is the number of atom types in the vocabulary.

To further improve training stability and enhance generalization, we also perturb the protein atoms by adding small Gaussian noise:
\begin{equation}
    \tilde{\mathbf{x}}_P = \mathbf{x}_P + \boldsymbol{\epsilon}, \quad \boldsymbol{\epsilon} \sim \mathcal{N}(\mathbf{0}, 0.1^2 \mathbf{I}),
\end{equation}
where $\mathbf{x}_P$ denotes the original 3D coordinates of protein atoms. This perturbation acts as a regularizer and prevents overfitting to rigid binding site conformations.

During training, the model observes the context atoms $\mathbf{M}_{\text{ctx}}$ and the noisy target $(\mathbf{x}_t, \mathbf{v}_t)$, together with the perturbed protein input $\tilde{\mathcal{P}}$, and is optimized to recover the clean target $(\mathbf{x}_0, \mathbf{v}_0)$ by learning the reverse denoising posteriors:
\begin{align}
    p_\theta(\mathbf{x}_0 \mid \mathbf{x}_t, \mathbf{M}_{\text{ctx}}, \tilde{\mathcal{P}}), \quad
    p_\theta(\mathbf{v}_0 \mid \mathbf{v}_t, \mathbf{M}_{\text{ctx}}, \tilde{\mathcal{P}}).
\end{align}

This formulation enables conditional molecule generation under a unified framework, where different generation tasks are cast as masked substructure denoising problems, conditioned on their respective chemical and structural contexts.

\subsection{Initialization Strategy in the Inference Phase}

A critical consideration during inference is the selection of the initial position for the ligand atoms in the diffusion trajectory. To ensure consistency across tasks and to avoid introducing spurious position-dependent biases, we adopt task-specific initialization strategies aligned with the underlying equivariance principles.

For \textit{lead optimization} tasks, where a known substructure is available, the initial molecule $\mathbf{m}_t$ is centered at the geometric center of the reference ligand to guide generation around its binding pose:
\begin{equation}
    \mathbf{x}_{t}^{(i)} \sim \mathcal{N}(\text{CoM}(\mathcal{L}), \sigma^2 \mathbf{I}),
\end{equation}
where $\text{CoM}(\mathcal{L}) = \frac{1}{N_L} \sum_{j=1}^{N_L} \mathbf{x}_L^{(j)}$ denotes the center of mass of the input substructure, and $N_L$ is the number of atoms in the ligand.

In contrast, for \textit{de novo} generation tasks without any reference substructure, the molecule is initialized around the center of the protein binding pocket:
\begin{equation}
    \mathbf{x}_{t}^{(i)} \sim \mathcal{N}(\text{CoM}(\mathcal{P}), \sigma^2 \mathbf{I}),
\end{equation}
where $\text{CoM}(\mathcal{P}) = \frac{1}{N_P} \sum_{j=1}^{N_P} \mathbf{x}_P^{(j)}$ denotes the center of mass of the protein pocket, and $N_P$ is the number of atoms in the pocket.

In both cases, the entire molecule is further translated to its own geometric center to preserve SE(3) invariance during generation.

\subsection{Task-Aware Dataset Construction}

We construct our training dataset based on the CrossDocked dataset \cite{crossdock}, which contains protein-ligand binding pairs covering a diverse chemical and structural space. Each molecule is assigned to one or more tasks (e.g., \textsc{Fragment}, \textsc{Linker}, \textsc{Scaffold}, \textsc{SideChain}) based on its structural context and suitability for substructure masking. In the single-task setting, each model is trained solely on the molecules annotated for that specific task. In contrast, our multi-task framework employs a dynamic task sampling strategy: during training, a task $t$ is sampled for each molecule $\mathbf{m}$ at every iteration, and a corresponding masking transformation $\mathcal{F}_t(\mathbf{m})$ is applied. This design allows the same molecule to participate in different generative contexts across training, thereby enhancing data efficiency and regularization.

\begin{table}[h]
\centering
\setlength{\tabcolsep}{4pt}    
\renewcommand{\arraystretch}{0.9}
\caption{Train split overlap between molecular generation tasks.}
\label{tab:train_overlap}
\begin{tabular}{l l c c c}
\toprule
Task A & Task B & Task A TrainSet & Task B TrainSet & Overlap \\
\midrule
\textsc{Frag}     & \textsc{Linker}     & 101{,}040 & 87{,}788  & 80{,}486 \\
\textsc{Frag}     & \textsc{Scaffold}   & 101{,}040 & 114{,}387 & 75{,}766 \\
\textsc{Linker}   & \textsc{Scaffold}   & 104{,}788 & 114{,}387 & 63{,}321 \\
\textsc{Scaffold} & \textsc{SideChain}  & 114{,}387 & 114{,}387 & 0 \\
\bottomrule
\end{tabular}
\end{table}

Table \ref{tab:train_overlap} shows the distributional overlap among training subsets for different tasks. There is substantial intersection between \textsc{Fragment}, \textsc{Linker}, and \textsc{Scaffold}, suggesting shared structural priors across these generation modes. Interestingly, the \textsc{SideChain} task is built on the same molecule set as \textsc{Scaffold}, but focuses on replacing side chains rather than decorating core scaffolds. This underscores the core motivation of our multi-task framework: enabling a unified model to learn diverse generation behaviors from structurally shared but semantically distinct supervision.

\section{Related Work}

\subsection*{Single-task 3D Generative Models} 
Early 3D molecular generators remain largely siloed. Diffusion models such as EDM \cite{edm} and GeoDiff \cite{geodiff} must be retrained for each task, while bespoke variants—DeLinker \cite{DeLinker}, MolGrow \cite{molgrow}, ScaffoldVAE \cite{ScaffoldVAE}—encode hand-tuned geometric rules but stay confined to their niches. Even Proformer \cite{proformer}, tailored to PROTAC linker design, relies on reinforcement learning for a single objective. Collectively, these pipelines generalize poorly and miss shared spatial–chemical regularities, underscoring the need for a unified, geometry-aware generative framework.

\subsection*{SMILES Pretraining and Transfer Learning} 
Large-scale SMILES pre-training has produced capable sequence models—ChemBERTa \cite{chemberta}, MolBERT \cite{molbert}, MolT5 \cite{molt5} for representation learning and MolGPT \cite{molgpt}, SAFE-GPT \cite{SAFE-GPT} for generation—that treat molecules as text, using masked-token or autoregressive objectives to support property prediction, retrosynthesis, and library design. Yet the 1-D string abstraction discards three-dimensional geometry: steric interactions, conformational strain, and shape complementarity are invisible, and protein context is absent, so target engagement must be approximated with external heuristics. Accordingly, SMILES models perform well on distributional or property-driven tasks but falter on spatially sensitive objectives such as linker closure or structure-based ligand design.

\subsection*{Toward Unified 3D Frameworks}
Unified 3D-aware frameworks remain aspirational. Uni-Mol \cite{unimol} and Uni-MoMo \cite{UniMoMo} pretrain geometric encoders with contrastive or denoising objectives, then append task-specific heads, while graph translators extend autoregressive decoders to multiple edit operations but still embed bespoke conditioning logic. Mol-Instructions \cite{molinstruction} borrows instruction tuning from NLP, allowing task switches through natural-language prompts, yet operates on SMILES and thus lacks native spatial reasoning. Consequently, these approaches struggle with geometry-dependent tasks—fragment growth, linker insertion, scaffold hopping—particularly in zero-shot scenarios that require precise ligand–receptor co-design. No existing model achieves true compositional generalization across structure-conditioned objectives without auxiliary labels or handcrafted heuristics, underscoring the need for an instruction-following 3D generative framework.

\subsection*{Position of the Present Work}
Our approach, Mask Once, Design All, differs from prior efforts in two key aspects. First, it unifies the four canonical editing tasks under a \emph{single} masked reconstruction objective applied directly to 3D coordinates, dispensing with task labels and handcrafted constraints. Second, the resulting model exhibits strong zero-shot transfer and creative adaptability, properties not reported by single-task specialists or label-conditioned multi-task baselines. By demonstrating that a carefully designed multi-task objective can outperform a collection of task-specific pipelines, this work establishes a new direction for general-purpose molecular generators.

\bibliographystyle{IEEEtran}
\bibliography{references}

\end{document}